\documentclass[10pt]{article}

\usepackage[superscript,sort]{cite}

\usepackage{ifthen,ifpdf}
\ifpdf
	\usepackage[pdftex]{hyperref}	\hypersetup{colorlinks=true,linkcolor=black,citecolor=black,urlcolor=blue,backref=page,bookmarks=true,breaklinks=true,plainpages=false }
	\usepackage[pdftex]{graphicx}
	\usepackage[usenames,pdftex]{color}
\else
	\usepackage[colorlinks=true,breaklinks=true,plainpages=false]{hyperref}
	\usepackage{graphicx}
	\usepackage[usenames]{color}
\fi

\usepackage{amsthm,amscd,amsxtra,amsfonts,amsmath,amssymb,multirow}
\usepackage{wrapfig}
\usepackage[footnotesize]{caption}
\usepackage[tiny,compact]{titlesec}
\usepackage[textwidth=0.8in,textsize=footnotesize]{todonotes}
\usepackage{algorithm,algorithmic,extarrows}

\begin{document}

\title{Sequence-based Multiscale Model (SeqMM) for High-throughput chromosome conformation capture (Hi-C) data analysis}

\author{
Kelin Xia$^{1,2}$ \footnote{ Address correspondences  to Kelin Xia. E-mail:xiakelin@ntu.edu.sg}\\
$^1$Division of Mathematical Sciences, School of Physical and Mathematical Sciences, \\
Nanyang Technological University, Singapore 637371\\
$^2$School of Biological Sciences \\
Nanyang Technological University, Singapore 637371\\
}

\date{\today}
\maketitle

\begin{abstract}
In this paper, I introduce a Sequence-based Multiscale Model (SeqMM) for the biomolecular data analysis. Biomolecular data is fundamentally different from the general point cloud data due to its unique sequential information. Traditional clustering methods, including K-means, hierarchical clustering, graph model, network model, spectral graph models, modularity, etc, focus more on the Euclidean distance information and tend to ignore sequential details. Therefore, they should not be applied directly to biomolecular data when sequential information matters. In my SeqMM, I introduce a sequence scale parameter to systematically remove the sequentially long-range interactions and generate a sequence-based multiscale model. With the combination of spectral graph method, I reveal the essential difference between the global scale models and local scale ones in structure clustering, i.e., different optimization on Euclidean (or spatial) distances and sequential (or genomic) distances. More specifically, clusters from global scale models optimize Euclidean distance relations. Local scale models, on the other hand, result in clusters that optimize the genomic distance relations. For a biomolecular data, Euclidean distances and sequential distances are two independent variables, which can never be optimized simultaneously in data clustering. However, sequence scale in my SeqMM can work as a tuning parameter that balances these two variables and deliver different clusterings based on my purposes. A DNA example is considered to illustrate my SeqMM in data clustering.  Further, my SeqMM is used to explore the hierarchial structures of chromosomes. To validate my model, I study genomic compartments and topological associated domains (TADs). I find that in global scale, the Fiedler vector from my SeqMM bears a great similarity with the principal vector from principal component analysis, and can be used to study genomic compartments. In TAD analysis, I find that TADs evaluated from different scales are not consistent and vary a lot. Particularly when the sequence scale is small, the calculated TAD boundaries are dramatically different. Even for regions with high contact frequencies, TAD regions show no obvious consistence. However, when the scale value increases further, although TADs are still quite different, TAD boundaries in these high contact frequency regions become more and more consistent. Finally, I find that for a fixed local scale, my method can deliver very robust TAD boundaries in different cluster numbers.
\end{abstract}

\newpage

\section*{Introduction}

Chromosome, the physical realization of genetic information, is one of the most complex and important cellular entities\cite{Bolzer:2005three,Hou2012:gene,Duan:2010three,sexton:2012three,Tanizawa:2010mapping,Zhang:2013chromatin,Sanyal:2011chromatin}. Over the past few decades, the significance of its three-dimensional architecture for essential biological functions, such as DNA replication, transcription, repair of DNA damage, chromosome translocation, etc, has gradually been realized\cite{Cavalli:2013functional,Chen2015:functional,Le:2014distinct,Pope:2014topologically}. Chromosome conformations are found to be deeply involved in the emergence of epigenetic organization, regulation of genome function and epigenetic inheritance of various cell states\cite{Cavalli:2013functional}. A thorough understanding of chromosome three-dimensional structure is fundamental to the decryption and interpretation of genetic information, and gradually becomes the most important topic in genomic and epigenetic research. However, chromosome structure is not only hierarchial and complicated, but also highly dynamic and variable between different cells. Chromosome folding mechanism remains largely elusive. Historically, various techniques have been proposed to explore the chromosome configuration \cite{Bonev2016:organization}, such as microscopy-based approaches and fluorescent in situ hybridization (FISH). Microscopy-based techniques and FISH are developed for the visualization of the chromosome three-dimensional structure. Currently, they can be used to study the spatial relations of a small number of genetic loci. More recently, Chromosome conformation capture (3C) technique \cite{Dekker:2002capturing,De2012:decade} and its derived methods, including chromosome conformation capture-on-chip (4C)\cite{Simonis:2006nuclear,Zhao:2006circular}, chromosome conformation capture carbon copy (5C)\cite{Dostie:2006chromosome} and High-throughput chromosome conformation capture (Hi-C)\cite{Lieberman:2009comprehensive}, have been developed and began to uncovered general features of genome organization\cite{Lieberman:2009comprehensive,Dixon2012:topological,Nora2012:spatial,Dryden:2014unbiased,Dixon:2015chromatin,Schoenfelder:2015pluripotent,Bonev2016:organization,Schmitt:2016genome,Nagano:2013single}.

The standard C-techniques (3C, 4C, 5C, Hi-C) involves several steps: crosslinking chromatin, digestion by restriction enzymes, proximity ligation, DNA purification and paired-end sequencing\cite{De2012:decade,Lieberman:2009comprehensive,Schmitt:2016genome}. In chromatin crosslinking, formaldehyde is used to fix the chromatin structure by crosslinking the chromatin proteins to their associated DNA. The crosslinked DNA will then be cut into chromatin fragments at enzyme-specific recognition motifs. This process is done by restriction enzymes. For example, endonucleases such as DNase I \cite{Ma:2015fine} will fragment DNA at sites of open chromatin. Micrococcal nuclease (MNase) will cut chromatin in histone linker sequences. After that, all the digested DNA fragments will be put into the diluted condition to facilitate the proximity re-ligation, which happens at ends of the juxtaposed DNA fragments. To enrich for ligation junctions and reduce unligated DNA fragments, before the proximity re-ligation, biotin-conjugated nucleotides are attached to the enzyme digestion sites at $5'$ overhangs. Further, biotin-labelled DNA fragments are isolated by affinity purification. Finally,
the frequency of ligation between two genomic loci is then assessed using PCR or direct DNA sequencing. The generated genome-wide chromatin contact maps give a representation of the chromatin structure. It should be noticed that C-techniques are usually employed on a large population of cells, so chromosome contact frequencies are average values over the whole population. However, single-cell approaches have also been developed \cite{Nagano:2013single}.


With the increasing availability of genome-wide C-techniques \cite{Dekker:2002capturing,Simonis:2006nuclear,Zhao:2006circular,Fullwood:2009oestrogen, De2012:decade,Lieberman:2009comprehensive,Dixon2012:topological,Nora2012:spatial,Jin2013:high,Bonev2016:organization,Schmitt:2016genome,Nagano:2013single} and C-data for multiple species and tissues, researchers begin to understand more about the folding and organization principles of chromosomes. In general, the structure of mammalian chromosomes can be explored from several different scales\cite{Bonev2016:organization}, including nucleosome, chromatin fiber, chromatin loops\cite{Rao:2014A3D}, topological associated domain (TAD)\cite{Dixon2012:topological,Nora2012:spatial}, genomic compartment\cite{Lieberman:2009comprehensive}, chromosome territory\cite{Lieberman:2009comprehensive}, etc. A nucleosome is a basic building block for chromatin organization. They interact with each other to form the 30 nm chromatin fibres with solenoid or zigzag shape. When I zoom out to a large scale, chromatin loop emerges. Chromatin loop is formed when cis-regulatory element, such as enhancers, are folded into close spatial proximity with its target promoter. This long-range chromatin contacts are important to tRNA genes, centromeres, early origins of replication, transcription factories for regulation of gene expression. Recent studies on Hi-C data have demonstrated the existence of an even larger scale structure known as topologically associating
domains (TADs)\cite{Dixon2012:topological,Nora2012:spatial}. TADs are chromosome components that are about 200 kilobases to 2 megabases. They are originally found as the contiguous square regions along the diagonal of Hi-C maps with large contact values. More importantly, TADs are very consistent between different cell types and species and their spatial distributions are highly correlated with many genomic features such as histone modifications, coordinated gene expression, association with the lamina and DNA replication timing. Another finding from Hi-C data is the genomic compartment\cite{Lieberman:2009comprehensive}. Through the principle component analysis, two types of compartments, i.e., A and B, are identified. More specifically, compartment B is more densely packed with higher interaction frequency. On the contrary, compartment A is chromosome regions that are more open and accessible. It strongly correlates with the gene loci and higher gene expression. More recently, analysis on the Hi-C data with 1kb resolution indicate the existence of six different subcompartments\cite{Rao:2014A3D}. It should be noticed that compartments are at a larger scale than TADs. Normally, compartment structures are not conserved between different types of cells. Finally, I arrive at the largest scales and the corresponding structure is known as the chromosome territory\cite{Lieberman:2009comprehensive}. Basically, each chromosome is an individual territory that seldom intermixes with others. The interactions between loci on the same chromosome territory are much more frequent than interactions between different chromosomes. Essentially, the chromosome is a multi-scale structure. This multi-level architecture is regulated and exploited by a variety of components such as transcription factors, architectural proteins and non-coding RNAs in order to coordinate gene expression and cell fate.


Even though Hi-C technique is a powerful tool to study the chromosome structure, it still faces challenges\cite{Yaffe:2011probabilistic,Imakaev:2012iterative, Ay:2014statistical,Witten:2012assessment}. Firstly, biases that complicate
the interpretation of observed contact frequencies can be introduced from procedures including crosslinking, chromatin fragmentation, biotin-labelling and religation. Particularly, systematic biases can substantially affect the Hi-C experimental results comes from three major sources\cite{Yaffe:2011probabilistic}, including distance-between restriction sites, the GC content and sequence mappability. Accounting for these biases is the first and most important step in C-data analysis. Various methods have been proposed to remove these biases, including HiCNorm\cite{Hu:2012hicnorm}, vanilla coverage\cite{Rao:2014A3D,Lieberman:2009comprehensive}, iterative correction and eigenvector decomposition (ICE)\cite{Imakaev:2012iterative}, Matrix-balancing method\cite{Knight:2013fast}, etc. Secondly, Hi-C maps generally do not represent chromosome conformation from a single cell, instead they are population-average map derived from hundreds of millions of cells. Single-cell Hi-C experiments show that contact maps of individual cells are highly variable\cite{Nagano:2013single}. The relationship between Hi-C contact frequencies and locus spatial distances is highly complicated. Sometimes, Hi-C contact frequencies may increase even when two loci move away from each other\cite{Imakaev2015modeling}. Reconstruction of the 3D structure of the genome from the C-data has been intensively studied\cite{Bau:2011three,Hu:2013bayesian,Zhang:20133d,Segal:2014reproducibility,Lesne:20143d,Zhang:2015topology,Imakaev2015modeling}.

Based on C-data, various algorithms and models are proposed to study the hierarchical structure of chromosome\cite{Lieberman:2009comprehensive,Dixon2012:topological,Filippova2014:identification,levy:2014two,Bau:2011three,Hu:2013bayesian,Zhang:20133d,Segal:2014reproducibility,Lesne:20143d,Zhang:2015topology,Imakaev2015modeling}. As mentioned above, one of the most important chromosome structure is genomic compartment\cite{Lieberman:2009comprehensive}. Computationally, genomic compartments are first founded from principal component analysis. Later, clustering approaches including hierarchical, k-means, graph and spectral graph are used to identify conserved chromosomal modules\cite{Siahpirani:2016multitask,Boulos:2013revealing,Wang:2013topological,Libbrecht:2015joint,Botta:2010intra}. Since TADs are essential to the understanding of relationship between chromosome structure and gene transcription, developing efficient algorithms for detecting TADs is another important topic in C-data analysis. Computationally, hidden Markov model (HMM) is the first method to identify TADs\cite{Dixon2012:topological}. In this model, a directionality index is calculated based on the contacts located 2Mb upstream and downstream of the current loci and used to capture the sharp transitions at TADs boundaries. After that, arrowhead algorithm with a ``corner score" is proposed\cite{Rao:2014A3D}. This special score indicates the likelihood of each locus to be TADs boundaries and can be efficiently evaluated by using dynamic programming. Meanwhile, resolution parameter is considered to identify TADs at various scales. This algorithm has been incorporated into the software Armatus\cite{Filippova2014:identification}. Further, a block-wise segmentation model called HiCseg \cite{levy:2014two} is proposed. This method reduces the problem of maximizing the likelihood with respect to the block boundaries into a 1D segmentation problem, and then employ the standard dynamic programming. 
More recently, a spectral graph theory based model is developed for the identification of TAD\cite{Chen2016:spectral}. In this model, Laplacian based graph segmentation is applied iteratively to obtain TADs at the given compactness level. All the above mentioned methods can be roughly divided into two categories, optimization method (HMM or dynamic programming) based local models and graph representation based global models. In local models, TAD indicators, including directionality index, corner score, likelihood of TAD boundaries, block-segmentation, are all evaluated locally within a certain region. In global models, TAD indicators, including eigenvectors, within-cluster variance, cluster distances, etc, are all evaluated globally in the whole domain. In this paper, a Sequence-based Multiscale Model (SeqMM) is introduced to put local models and global models in the same footing. I find that the essential difference between the global scale models and local scale ones in structure clustering lies in the optimization of Euclidean (or spatial) distances or sequential (or genomic) distances, which are two faces of biomolecular data.

Biomolecular data is fundamentally different from the general point cloud data due to its unique sequential information. Traditional clustering methods, including K-means, hierarchical clustering, spectral graph models, etc, focus more on the Euclidean distance information and tend to ignore sequential details. Therefore they can not be applied directly to biomolecular data when sequential information matters. To overcome this problem, I introduce a sequence scale parameter to systematically remove sequentially long-range interactions in my SeqMM. In this way, the resulting matrixes naturally provide a representation of the hierarchical biomolecular structure. Further, with the combination of spectral graph method, I point out that in structure clustering, the global scale models and local scale ones have different emphasis on Euclidean distances and sequential distances. More specifically, clusters from global scale models optimize Euclidean distance relations. Clusters from local scale models optimize genomic distance relations. For the clustering of biomolecular data, these two independent variables, Euclidean distances and sequential distances, can never be optimized simultaneously. And the sequence scale in my SeqMM can work as a tuning parameter that balances these two variables and deliver different clusterings based on my purposes. I demonstrate different domain decompositions in my SeqMM using a DNA example. After that, my SeqMM is used to explore the hierarchial structures of chromosomes, particularly, I study genomic compartments and topological associated domains (TADs). It is found that in global scale, the Fiedler vector from my SeqMM bears a great similarity with the principal vector from principal component analysis, and can be used to study genomic compartments. In TAD analysis, I find that different sequence scales give different TAD boundaries. When sequence scale is small, a littler change of the scale will result in TAD boundaries that are dramatically different. No obvious consistence can be seen even for regions with high contact frequencies. However, when the scale values increase further, although TADs are still quite different, TAD boundaries in these high contact frequency regions become more and more consistent. Finally, I find that for a fixed local scale, my method can deliver very robust TAD boundaries in different cluster numbers.




\section*{Model and results}\label{sec:theory}

\begin{figure}
\begin{center}
\begin{tabular}{c}
\includegraphics[width=0.8\textwidth]{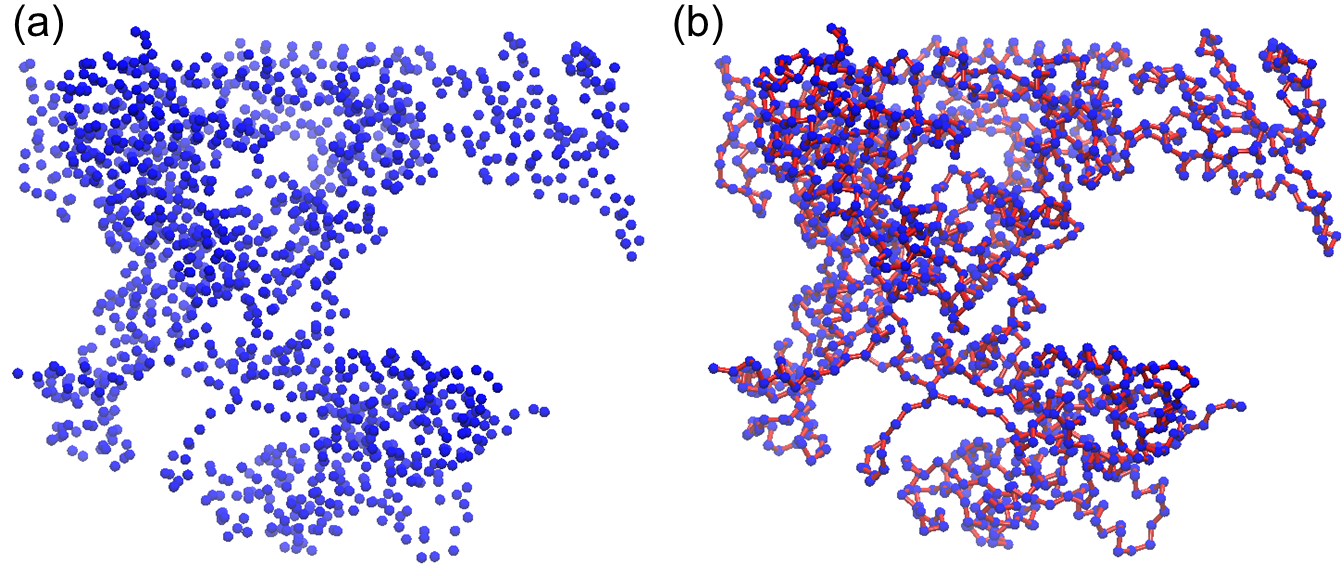}
\end{tabular}
\end{center}
\caption{ Difference between a point cloud data and a biomolecular structure data from protein 5U5Q. In general, biomolecular structure data is a sequence-based point cloud data. The introduction of sequential information has brought in one more condition.
}
\label{fig:point_cloud_data}
\end{figure}

One of the major features of biological sciences in the $21^{st}$ century is its transition from an empirical, qualitative and phenomenological discipline to a comprehensive, quantitative and predictive one. Revolutionary opportunities have arisen for data-driven advances in biological research. However, these opportunities are associated with severe challenges: the emergence of complexity in self-organizing biological systems and their associated massive datasets poses fabulous challenges to their quantitative description and prediction. Among these challenges is the understanding of chromosome hierarchical structure and its folding mechanism. Chromosome three-dimensional architecture is found to play an important role in essential biological functions such as DNA replication, transcription, repair of DNA damage, chromosome translocation, etc. Therefore, the decryption and interpretation of chromosomal genetic information from Hi-C data has gradually becomes more and more significant in genomic and epigenetic research. In this section, I will introduce a Sequence-based multiscale model for biomolecular structure data analysis. My model has revealed the fundamental difference in of various scales models in biomolecular structure data clustering. To illustrate the basic idea, I use a DNA structure as an example. Further, my SeqMM is employed for Hi-C data analysis. My model is able to repeat the genomic compartment results from principal component analysis. TADs are studied with my local scale models. Some basic results have been presented.

\subsection*{Sequence-based multiscale modeling}\label{sec:seqmm}
\begin{figure}
\begin{center}
\begin{tabular}{c}
\includegraphics[width=0.8\textwidth]{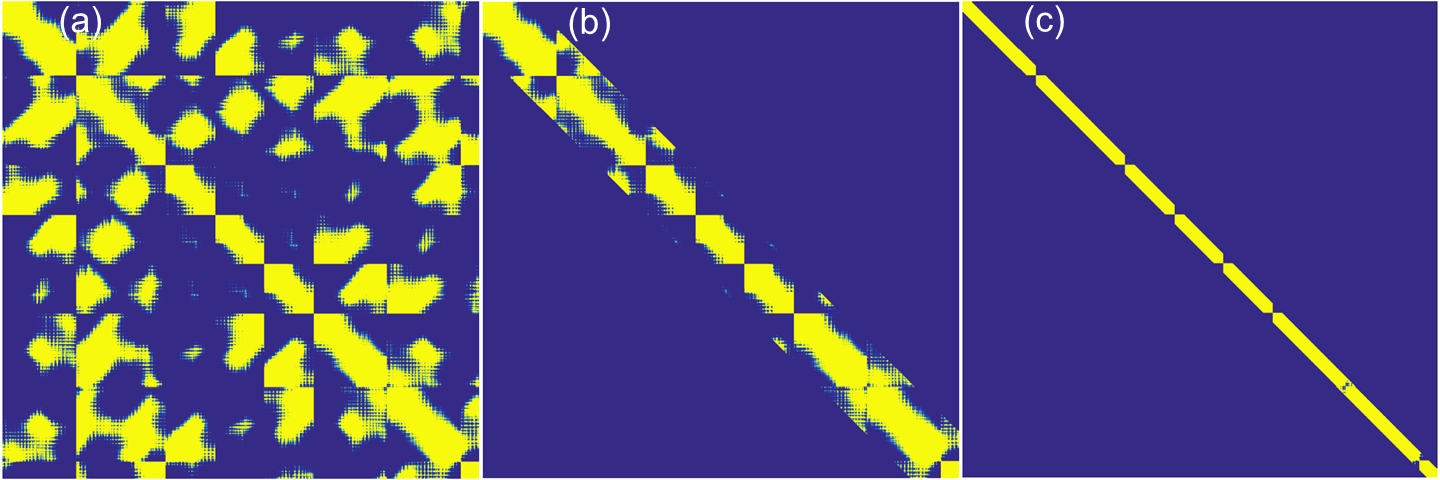}
\end{tabular}
\end{center}
\caption{Three different SeqMM weight matrixes derived from protein 5I6D. The values of sequence scale parameter $\eta$ are $1.0$, $0.1$ and $0.02$ in subfigure {\bf (a)}, {\bf (b)} and {\bf (c)}, respectively.
}
\label{fig:dna_matrix}
\end{figure}

\begin{figure}
\begin{center}
\begin{tabular}{c}
\includegraphics[width=0.8\textwidth]{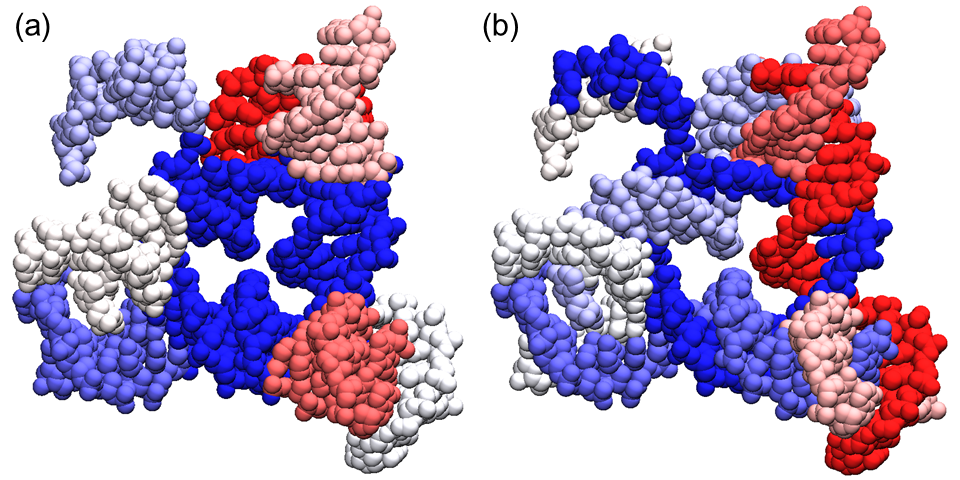}
\end{tabular}
\end{center}
\caption{ Clustering of DNA 5I6D into eight regions using two different scale values $1.0$ {\bf (a)} and $0.1$ {\bf (b)}. The clustering is done by the combination of the spectral graph model and K-means algorithm. {\bf (a)} The global scale model. Atoms in each cluster are spatially more close to each other. {\bf (b)} The local scale model. Atoms in each cluster are sequentially more close to each other. In fact, each cluster represents an individual chain in DNA.  From the comparison, it can be seen that clusters from global scale model optimizes Euclidean distance relations. Clusters from local scale model optimizes sequence distance relations.
}
\label{fig:dna_clustering}
\end{figure}

\begin{figure}
\begin{center}
\begin{tabular}{c}
\includegraphics[width=0.5\textwidth]{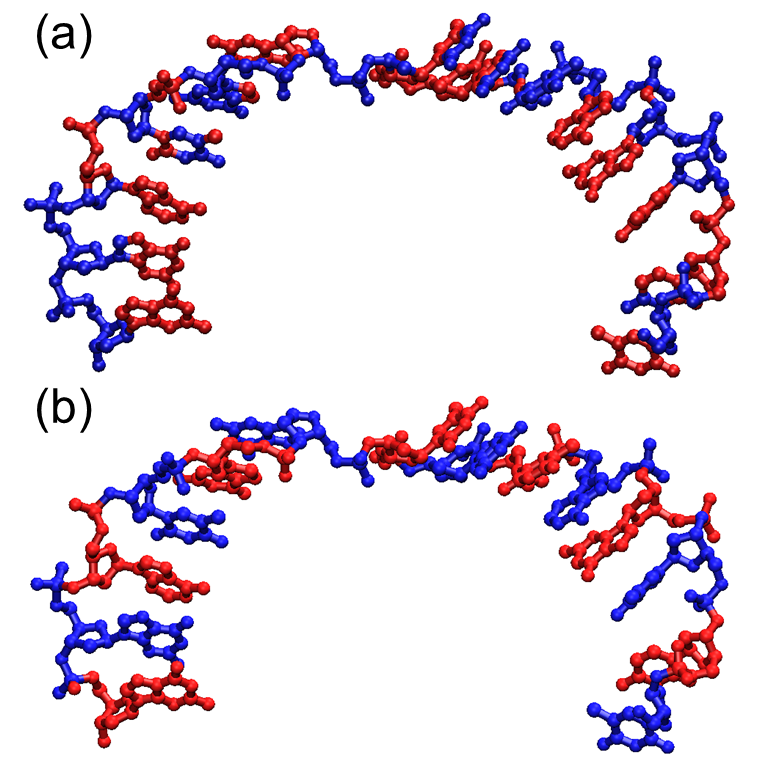}
\end{tabular}
\end{center}
\caption{The clustering of DNA 5I6D into 128 clusters clusters using two different scale values $1.0$ {\bf (a)} and $0.02$ {\bf (b)}. The clustering is done by the combination of the spectral graph model and K-means algorithm. Again, I can find that for global scale model in {\bf (a)}, atoms in each cluster are spatially more close to each other.For local scale model in {\bf (b)}, atoms in each cluster are sequentially more close to each other. In fact, each cluster represents an individual nucleic acid.  The results confirm my finding that clusters from global scale model optimizes Euclidean distance relations. Clusters from local scale model optimizes sequence distance relations.
}
\label{fig:dna_residue}
\end{figure}

Biomolecular structure data is essentially different from the general point cloud data (PCD), as it incorporates a unique sequential information. It is well-known that proteins, DNAs, RNAs, and their complexes like chromatin, are all made from one or several chains. The sequential information is then embedded into these polymer chains, which will coil into special three-dimensional structures to fulfill certain biological functions. I have demonstrated the difference between PCD and biomolecular structure data in Figure \ref{fig:point_cloud_data}. Simply speaking, biomolecular structure data can be regarded as sequence-based PCD. However, the sequential information, which is of great importance in hierarchical structure analysis, is rarely considered in PCD analysis. This is due to the reason that structural PCD is usually a discrete representation of certain underlying geometries, manifolds, high-dimensional structures, or complicated subjects. PCD itself is just the aggregation of isolated points with no significant features. However, the topology and connectivity information of the hidden structure can be analyzed by transforming PCD into special graph or network models. Further, based on these models, classification or clustering methods, including K-means, hierarchical clustering, spectral clustering, modularity, graph centrality, network approaches, etc, are used to explore the properties of the associated structures. It should be noticed that this transformation is usually non-trivial. Geometric, physical, chemical and biological properties are usually considered in this process. Even though some of the sequential information is considered in atomic bonds, bond angles and dihedral angles, no special attention is paid to this unique biomolecular structure property. And the sequential information is by no means irrelevant. In this section I  will demonstrate that sequential information can play significant roles in structure analysis, particularly in structure classification and domain decomposition.

Before the formal introduction of my model, I need to clarify two essential distance definitions, i.e., Euclidean distance and sequential distance. Euclidean distance is just distance between two subjects spatially. In Hi-C data, Euclidean distance can be inversely related to contact frequency roughly\cite{Imakaev2015modeling}. Sequential distance is defined between two elements located on the a one-dimensional chain, even though it can be embedded in higher dimensional spaces. When each element is assigned with a sequence number, the sequential distance is just the difference between these two numbers and it is always an integer. In Hi-C data, sequential distance between two locu is their genomic distance. Even though graph and network based multiscale models and multiscale analysis are widely used biomolecular structure and function analysis\cite{KLXia:2014a,XFeng:2012a,KLXia:2014c,KLXia:2015c,KLXia:2013d,Xia:2015multiscale,KLXia:2015d}, the scales defined in these models are in terms of Euclidean distance. To be more specific, when I discuss atomic scale, residue scale, second structure scale, tertiary structure scale, etc, I am analyzing structural elements based on their sizes or relative Euclidean distances. Dramatically different from all the previous multiscale models, the scale defined my SeqMM is not based on the Euclidean distance but on the sequential distances. On the other hand, previous multiscale models resort to a spatial distance related scale parameter for the controlling of model scales. For example, if elements in a biomolecule are only allowed to interact with others within a certain cutoff distance, a multiscale model can be built by changing this distance. That is to say, if elements is only allowed to ``communicate" with very adjacent ones, a localized model is attained. If elements are free to talk with all the other atoms, it is a globalized model. State differently, by tuning the scale parameter, one can analyze biomolecular structures from various scales\cite{KLXia:2014a,XFeng:2013b,KLXia:2013d}.

If the scale parameter is based on the sequential distance, similar analysis can be applied to my SeqMM. To facilitate a better understanding, I consider a DNA example with PDB ID 5I6T. In this DNA structure, there are eight chains denoted from chain A to chain H. Atoms in each chain are arranged based on nucleic acid sequence and their positions in nucleic acids, following the basic PDB format. In this way, all atoms are organized sequentially. This DNA structure can be represented by a distance matrix or a weighted matrix. I consider the weight matrix representation, because its values are comparable to contact frequencies in Hi-C data. More specifically, weight values are inversely related with the Euclidean distances and are constructed by my rigidity functions \cite{Opron:2014} as following,
\begin{eqnarray}\label{eq:weight}
M=\{ M_{ij}=e^{-\left( r_{ij} /\eta \right)^2} \}, i=1,2,...,N; j=1,2,...N,
\end{eqnarray}
where $r_{ij}=\|{\bf r}_i-{\bf r}_j \|$ is the Euclidean distance between the $i$th and $j$th atoms and $N$ is the total number of atoms. The parameter $\eta$ is a scale parameter that controls the influence range of each atom and in this case I chose it to be 8 \AA~ based on my previous results\cite{Opron:2014b,Opron:2015communication,Nguyen:2016generalized}. In the weight function, if two atoms are closed to each other, their weight value will be large, just like the contact frequency in Hi-C data.

In my SeqMM, I use $N_b$ to denote the contact range along the sequence. An atom is only allowed to contact with other
atoms that are along the sequence and between $N_b$-th upstream and downstream atoms. Based on this contact range, a new weight matrix can be expressed as,
\begin{eqnarray}\label{eq:Seq matrix}
M_{ij}^{\rm Seq}=\begin{cases} \begin{array}{ll}
            M_{ij} &  |i-j|>0 ~{\rm and}~ |i-j|\leqslant N_b\\
			 0     &  |i-j|>N_b \\
            -\sum_{i \neq j}^N M_{ij}^{\rm Seq} &i=j.
	      \end{array}
\end{cases}
\end{eqnarray}
Mathematically, my SeqMM matrix in Eq. (\ref{eq:Seq matrix}) is a weight Laplacian matrix, which plays a very important role in graph representation and spectral graph models. More details of Laplacian matrix will be discussed later. In my SeqMM matrix, short-range interactions within $N_b$ distance are untouched while long-range interactions beyond $N_b$ distance are systematically removed. Geometrically, this means the new structure represented by my SeqMM matrixes preserves the structural components that formed by adjacent atoms, but destroys all structural parts supported by the long-range interactions.

I can systematically change the contact range to deliver a series of weight matrixes in different scales. To facilitate the comparison between different scales, I define a sequence scale parameter $r_b$ as,
\begin{eqnarray}\label{eq:ratio}
r_{b}=\frac{N_b}{N}.
\end{eqnarray}
The contact range can be represented by $N_b=r_b \times N$. I consider three different scale values $r_{b}=1.0, 0.1$ and $0.02$ in Figure \ref{fig:dna_matrix}. It can been seen that, with the decrease of scale value, atoms gradually lose the ability to communicate with others that are far way sequentially. Essentially, scale parameter controls the influence region along the sequence. Smaller scale value will result in a more localized model, while larger sacle value will induce a more globalized model. This will have a dramatic effect on the clustering of biomolecular structures.

To reveal this effect, I perform spectral clustering analysis on the above three weight matrixes simultaneously. Firstly, I decompose DNA structure into eight domains based on scale values $r_{b}=1.0$ and $0.1$. I find that these two weight matrixes result in dramatically different clustering patterns. Figure \ref{fig:dna_clustering} demonstrates the results. I paint van der Waals surfaces with different colors to represent different clusters. When $r_{b}=1.0$, the center regions of DNA are classified into a single domain, while extruding regions around it are decomposed into seven different clusters, as illustrated in Figure \ref{fig:dna_clustering} ({\bf a}). In contrast, when $r_{b}=0.1$, clusters are spatially more scattered. In fact, each cluster is exactly an independent chain of the DNA. I show the results in Figure \ref{fig:dna_clustering} ({\bf b}).
Secondly, I decompose DNA structure into 128 components based on scale value $r_{b}=1.0$ and $0.02$. It is nontrivial to denote these 128 clusters by different colors. So I only extract chain C and color even and odd numbers of clusters by red and blue colors, respectively.  The results are demonstrated in Figure \ref{fig:dna_residue}. It is found that when scale value $r_{b}=1.0$, nucleobases are not classified into the same cluster with deoxyribose and phosphate group, as illustrated in Figure \ref{fig:dna_residue} ({\bf a}). In fact, each nucleobase is paired with the complementary nucleobase to form an individual group. This differs greatly from the situation when $r_{b}=0.02$. It can been see from Figure \ref{fig:dna_residue} ({\bf b}) that with this localized representation, each nucleic acid itself is classified into one group. This happens because when my scale value is small, atoms are only allowed to interact with sequential closed neighbours rather spatially closed neighbours. State different, atoms can not ``see" the surrounding partners if their sequence numbers differ greatly. From these two cases, it can be found that when a large scale value is used, atoms with small Euclidean distances will be grouped together regardless of their sequence distances. When a small scale value is considered, atoms with small sequence distances will be grouped together regardless of their Euclidean distances. Essentially, my scale parameter $r_b$ provides a balance between Euclidean distance based clustering and sequence distance based clustering.

\begin{figure}
\begin{center}
\begin{tabular}{c}
\includegraphics[width=0.8\textwidth]{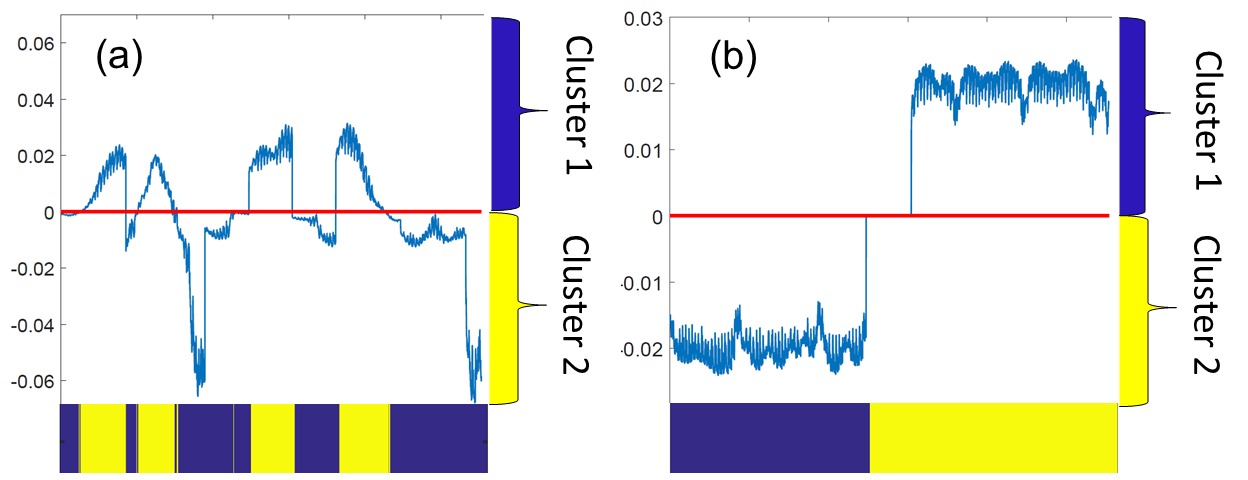}
\end{tabular}
\end{center}
\caption{ The comparison of the Fiedler vectors (eigenvectors from the second smallest eigenvalue) of the two SeqMM weighted Laplacian matrixes. {\bf (a)} Eigenvector derived from Laplacian matrix with $r_b=1.0$. {\bf (b)} Eigenvector derived from Laplacian matrix with $r_b=0.1$. The DNA can be decomposed into two clusters based on the sign of the eigenvector values. All atoms with positive eigenvector values are grouped into one cluster. The others fall into the second cluster. It can be seen that for global scale model, atoms are classified more randomly in terms of their sequential numbers. In comparison, local scale model are sequentially more ordered.
}
\label{fig:dna_spectral_analysis}
\end{figure}

This balance effect can be better understood by checking the Fiedler values and vectors from SeqMM matrixes. As stated above, the weighted matrix in Eq.(\ref{eq:Seq matrix}) is a weighted Laplacian matrix. Therefore, its second smallest eigenvalue, known as algebraic connectivity or Fiedler value, gives a representation of the overall connectivity of the underlying structure. Generally speaking, the magnitude of this value reflects how well connected the overall graph is. If this eigenvalue is greater than zero, the graph is a connected.  Fiedler value has been used in analysing the robustness and synchronizability of networks.
More interestingly, the corresponded eigenvector, known as Fiedler vector, gives an optimized partition and is widely used for graph decomposition. I employ the spectral clustering analysis on the weighted matrixes with $r_b=1.0$ and $0.1$. The calculated Fiedler values are 0.012 and 1.26e-8, respectively. Basically, small scale value result in less overall connectivity, just as I expected. The Fiedler vectors are depicted in Figure \ref{fig:dna_spectral_analysis} {\bf (a)} and {\bf (b)}. It should be noticed that each element or atom is associated with a unique value in Fiedler vector. In spectral clustering, I classify the element based on the sign of this unique value. As seen in Figure \ref{fig:dna_spectral_analysis}, atoms with positive Fiedler vector values are grouped together and represented by blue color. Atoms with non-positive values are classified into the other cluster represented by yellow color. Further, in global scale model, i.e., when $r_b=1.0$, atoms that are sequentially far away from each other can still communicate and be grouped into same cluster. This is largely due to the reason that global scale models emphasizes on the Euclidian distance but pays less attention to the sequential information. In comparison, atoms in the local scale model are only allowed to communicate with those sequentially neighboring to them. Thus the clustering is more sequentially ordered. To sum up, the scale parameter in my scale model provides a balance between global scale, which focuses more on Euclidian distance relationship, and local scale, which emphasizes more on sequential distance relationship.

Finally, it is worth mentioning that the folding mechanism of biomolecules and their complexes are always highly related to their hierarchical structures. Take protein as an example. The original polypeptide chain will form secondary structures first, then theses secondary structures interact to form tertiary structures, which later combine together to form quaternary structure. The process happens step by step and goes from local scales to global scales.  State differently, only after more local-scale structures are formed, global-scale structures can begin to emerge. More importantly, local-scale structures are created between elements that are sequentially close to each other. Global-scale structures are created under sequentially long-range interactions. Therefore if I represent the structures of biomolecules or their complexes in a matrix, the folding process and also the hierarchical structure can naturally be analyzed by systematically changing the sequence ratio. To be more specific, when sequence ratio is small, localized structures are captured. Topologically, the matrix band is very narrow. When I increase the sequence ratio, matrix band increase and more long-range interaction will be considered, meaning global scale information is now captured. State differently, different sequence ratios represent different scales in the data and my SeqMM provides a natural way to capture hierarchical structure properties.


\subsection*{Multiscale analysis of Hi-C data}\label{sec:multiscale_HiC}

The structure of mammalian chromosomes have multiple scales, ranging from chromosome territory\cite{Lieberman:2009comprehensive}, genomic compartment\cite{Lieberman:2009comprehensive}, topological associated domain and sub-TAD\cite{Dixon2012:topological,Nora2012:spatial}, chromatin loops\cite{Rao:2014A3D}, chromatin fiber, nucleosome, to atomic-scale structures such as minor and major grooves and nucleic acid pairs. Chromosome architecture is formed in a hierarchical manner. The most essential component is the nucleosome, which is made from nucleic acids. These highly dynamic nucleosomes can interact with each other to form 30-nm fibres. Spatially, these fibres can form loops to facilitate long-range interactions between promoters or enhancers and genes\cite{Schoenfelder:2015pluripotent,Sanyal:2012long}. Further, these loops can evolve into complicated motifs like TADs and sub-TADs. Biologically, architectural proteins, such as CTCT, cohesin, mediator, polycomb, etc, can help to establish and stabilize these loops, TADs and sub-TADs\cite{Bonev2016:organization,Phillips:2013architectural}. This multi-level architecture plays a fundamental role in various biological functions such as DNA replication, transcription, repair of DNA damage, chromosome translocation, etc.

As stated in the introduction part, C-techniques provide a unique way to understand the chromosome architecture structure. The values in C-data are contact frequencies between various loci in the chromosome. Even though the relationship between contact frequencies and locus spatial distances is highly complicated \cite{Imakaev2015modeling} and systematic biases can substantially affect the experimental results\cite{Yaffe:2011probabilistic}, including distance-between restriction sites, the GC content of trimmed ligation junctions and sequence uniqueness, C-data analysis has already reveal various interesting chromosome structure properties, like chromosome territory, genomic compartment, topological associated domain, etc.

In this section, I explore the potential application of SeqMM in Hi-C data analysis. Particularly, I discuss the use of SeqMM in genomic compartment analysis and TAD analysis. I find that my global scale model can deliver similar information of genomic compartments. More importantly, I find topological associated domains should be better represented by a local scale model instead of a global one. Using the spectral clustering and my sequence scale parameter, I naturally put the global scale models and local scale models in the same footing. I have also confirmed that my sequence scale provides a nature way to capture hierarchical structure properties. The details are given below.

\subsubsection*{General Hi-C data properties}\label{sec:compartment}

\begin{figure}
\begin{center}
\begin{tabular}{c}
\includegraphics[width=0.8\textwidth]{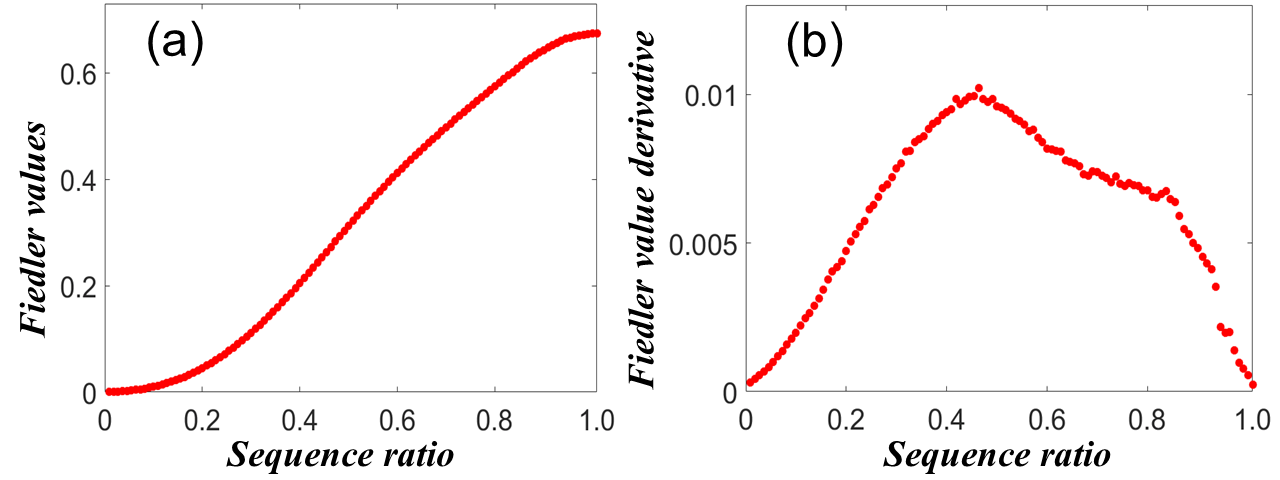}
\end{tabular}
\end{center}
\caption{Fiedler values from various Sequence-based scales of Hi-C data. I use GM06990 data with resolution 1Mb.
}
\label{fig:Fiedler_values}
\end{figure}

As stated in the introduction, the Hi-C data values are contact frequencies between different loci. Mathematically, it can be treated as a weighted adjacent matrix, which can be further transformed into a Laplacian matrix. With spectral graph algorithms, I can analyze the intrinsic connectivity and topology information of the Laplacian matrix, and derive Hi-C structure properties. The details of algorithms are discussed later. To explore the general behaviors of Hi-C data in various sequence ratios, I consider the 1Mb resolution GM06990 Hi-C data. Since the magnitude of Fiedler value reflects how well connected the overall graph is, I systematically calculate the Fiedler values for my SeqMM. Figure \ref{fig:Fiedler_values} demonstrates the results of Fiedler values and Fiedler value derivatives. It can be seen that Fielder value increases with the sequence ratio just as I expect. When the sequence ratio reaches at about 0.5, the increase rate arrives at its peaks. And after about 0.85, the increase rate decrease dramatically.

\begin{figure}
\begin{center}
\begin{tabular}{c}
\includegraphics[width=0.90\textwidth]{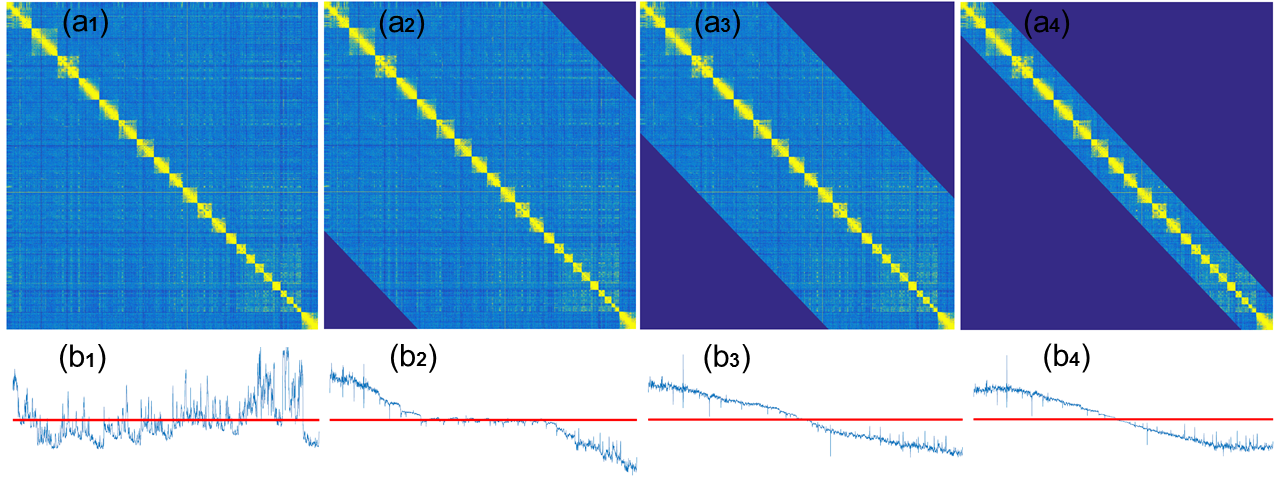}
\end{tabular}
\end{center}
\caption{ Sequence-based multiscale modeling of Hi-C data. I use GM06990 data with resolution 1Mb. The sequence ratio parameter $r_b=$ 1.0 , 0.7, 0.4 and 0.1 in subfigure {\bf (a)}, {\bf (b)}, {\bf (c)} and {\bf (d)}, respectively. The Fiedler vectors from spectral graph method are also demonstrated. It can be seen that with the decrease of the scale ratio, loci in each cluster becomes more organized sequentially. A clear transition from global connection to local interaction can be spotted.
}
\label{fig:multiscale_band}
\end{figure}

Further, I choose fmy different scale ratios $r_b=$ 1.0 , 0.7, 0.4 and 0.1 with Fiedler values 0.674, 0.503, 0.197 and 0.006, respectively, and calculate the correspondingly Fiedler vectors. The results are demonstrated in Figure \ref{fig:multiscale_band}. Fiedler vectors, as I expected, have similar behaviors as seen in the previous DNA case. When $r_b=1.0$, atoms that are sequentially far away from each other can still communicate and be grouped into same cluster. With the decrease of scale values, Fiedler vectors tend to have less fluctuations and atoms can only ``talk" to others that are sequentially closed to each other. Thus the clustering is more sequentially ordered. To sum up, the scale parameter provides a balance between global scale, which focuses more on Euclidian distance relationship, and local scale, which emphasizes more on sequential distance relationship. These results are highly consistent with my findings in DNA case. Therefore, I can expect that the same clustering properties found by my SeqMM in the DNA cases can also be seen in my Hi-C data.

\subsubsection*{Global scale based genomic compartment analysis}\label{sec:compartment}

\begin{figure}
\begin{center}
\begin{tabular}{c}
\includegraphics[width=0.90\textwidth]{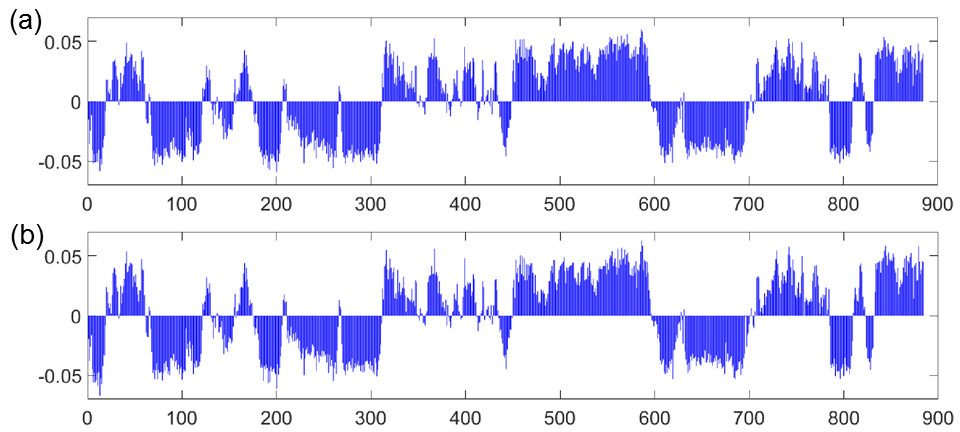}
\end{tabular}
\end{center}
\caption{The comparison of principal component from PCA and Fiedler vector from spectral graph theory. I use GM06990 chromosome 14 data with resolution 100kb. The global scale model with $r_b=1.0$ is used. {\bf (a)} Principal component (first eigenvector) results. {\bf (b)} Fiedler vector (second eigenvector) results. The two results are highly consistent with each other. This shows that spectral clustering in the global scale can capture the genomic compartment information very well.
}
\label{fig:global_scale}
\end{figure}

Eigenvectors derived from my spectral graph method can be used for domain decomposition. It is found that spectral clustering methods tend to outperform existing non-graph-based methods, producing higher quality clusters based on statistical enrichment of multiple one-dimensional regulatory genomic signals.
More interestingly, my Fiedler vector based genomic compartment analysis is comparable to the principal element from principal component analysis. To facilitate a better understanding, I consider the GM06990 chromosome 14 with resolution 100kb. This is a classic case used for genomic compartment analysis. However, before I employ the spectral graph analysis, I normalize the Hi-C data using the corresponding Toeplitz matrix, the same way as in papers. The normalized function is then transformed into my weight Laplacian matrix. Figure \ref{fig:global_scale} demonstrates the comparison between principal component {\bf (a)} and Fiedler vector {\bf (b)}. It can be seen that they bear great resemblance to each other. The same results can also be found in all the other chromosomes. It should be noticed that these results are consistent with previous findings in the paper.

\subsubsection*{Local scale based topological associated domain analysis}\label{sec:compartment}

Another very important finding in Hi-C data analysis is the topological associated domain. TADs are megabase-sized local chromatin interaction domains. They have loop structures and are highly stable across various cell types and conserved across species. TAD boundaries are found to be enriched with the protein CTCF, housekeeping genes, transfer RNAs and short interspersed element (SINE) retrotransposons. These components play important roles in establishing and supporting TADs and further the architectural structure of the chromosome. Due to the structural and functional importance of TADs, various algorithms are proposed for the identification of TADs as stated in the introduction part. Since it has already be pointed out that the problem of identification of TAD differs from the problem of clustering the graph represented by the Hi-C data, multiscale models and dynamic programming have been considered to meet this challenge. However, none of these methods provides a deep and mathematical understanding between clustering properties and sequence-based scales. In the following, I will reveal the relation between clustering properties and sequence ratios. 

\begin{figure}
\begin{center}
\begin{tabular}{c}
\includegraphics[width=0.4\textwidth]{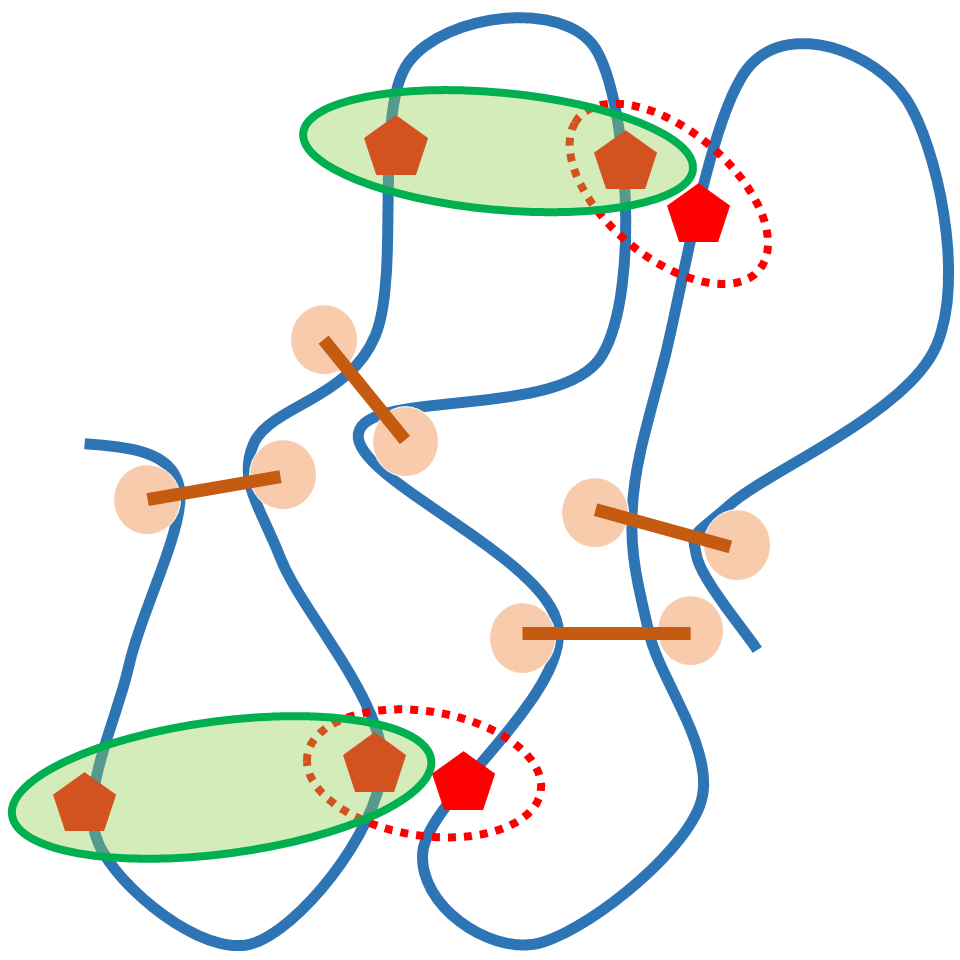}
\end{tabular}
\end{center}
\caption{ The illustrations of TAD analysis in various sequential scales. Each locus is represented as a red pentagon. Global scale models emphasize only the contact frequencies and will group loci that are closed into one cluster as indicated by the red dash circle. Using the right scale value, local scale models will be able to balance the sequential distance and contact frequencies, and group loci into biologically meaningful TADs as indicated by the blue sold circle.
}
\label{fig:illustration}
\end{figure}

The relation between clustering and sequential distances can be understood from a picture in Figure \ref{fig:illustration}. I have identified six loci represented by red pentagons. Geometrically, pairs of loci, that are enclosed by red circles, are much closer to each other than the pairs in green circles. Sequentially, However, pairs in green circles are more contiguous than the pairs in red circles. In this way, when I need to classify the data, I need to consider the balance between Euclidean distance and sequential distance (or genomic distance) just as I demonstrated in DNA case. Since TADs are localized compact regions, the sequential distances will have upper limit, usually around 5 Mb. In Figure \ref{fig:illustration}, it is obvious that even though loci enclosed by red dash circles are closer to each other geometrically, thus has higher contact frequencies, their sequential distance are too far away from each other and should not be grouped into the same TAD or sub-TAD. Again this can be comparable to the DNA clustering case, in which even though the nucleic base pairs are geometrically more favourable, they should not be classified into same cluster if sequential distance matters. Indeed, for the clustering of biomolecular data in different scales, I always need to balance between the Euclidean distance and the sequential distance.

However, to find a ``suitable" optimized sequence scale ratio for the identification of TADs is by no means trivial. Previously, in a hidden Markov model (HMM) based TADs method, a sequence range of 2 Mb upstream and downstream is used in the evaluation of its directionality index. For TAD software armatus, domain-length scaling factor, which is a sequence distance based parameter, is employed. In dynamic programming based approaches, including arrowhead and HiCseg, the sequential scale and size of TADs are optimized over the whole domain.

In my SeqMM based TAD analysis, a ``suitable" optimized scale value is also needed. To make things even more complicated, I do not know how many TADs are there in the data. State differently, I do not know the cluster number in my spectral clustering. Since the size of TADs are roughly around 1Mb, I define my cluster size $S_c$ in terms of the number of loci as $S_c=\frac{1000}{N_r}$, with $N_r$ the data resolution in the unite of kb. In this way, the cluster size $N_c$ can be estimated by
$$N_c=\frac{N}{S_c}=N\times \frac{N_r}{1000}.$$

\begin{figure}
\begin{center}
\begin{tabular}{c}
\includegraphics[width=0.90\textwidth]{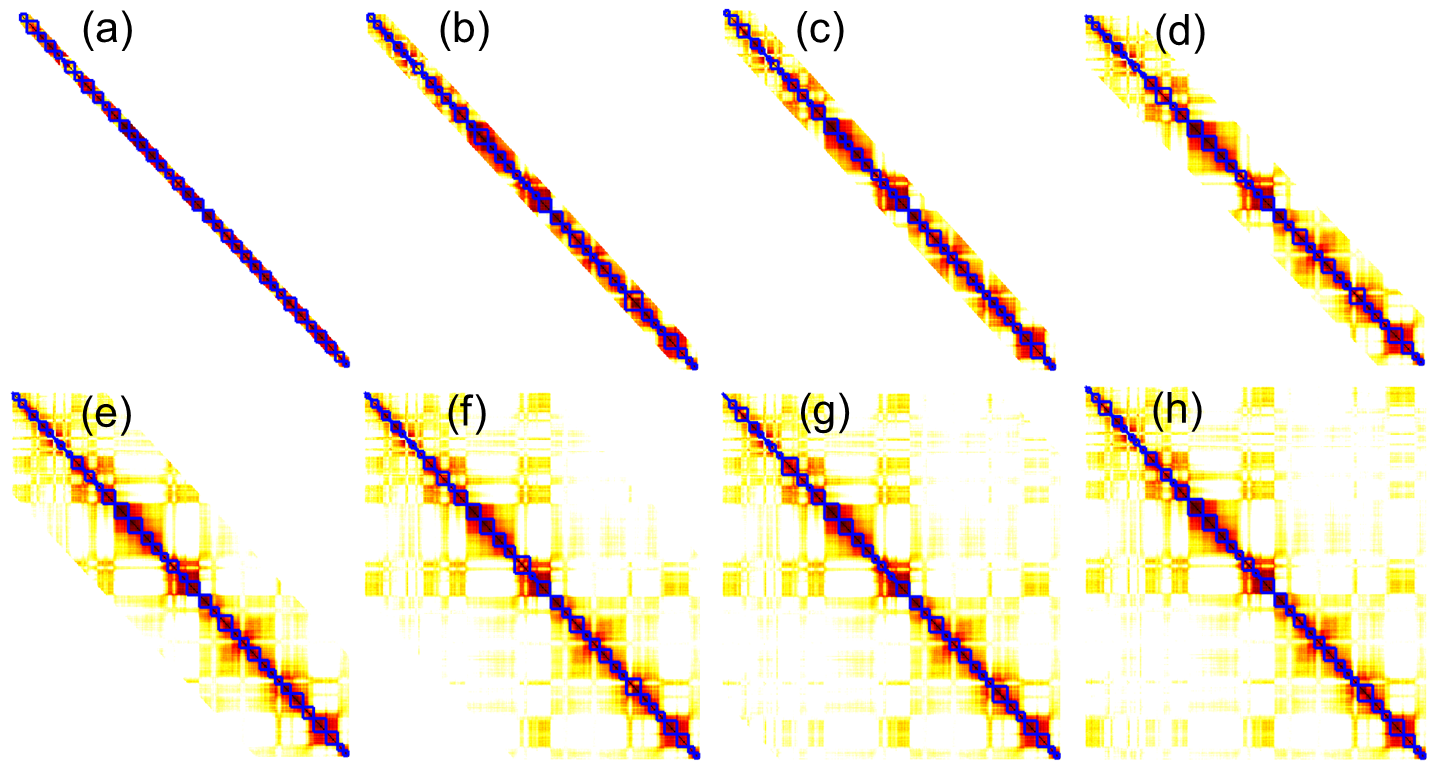}
\end{tabular}
\end{center}
\caption{ The calculated TADs from various band ratios. I use the IMR90 cell chromosome 22 with resolution 100kb. From {\bf (a)} to {\bf (h)}, the band ratios are $r_t=$1.0, 2.0, 3.0, 5.0, 10.0, 20.0, 30.0 and 40.0 (the whole matrix), respectively. The rectangles are identified TADs.
}
\label{fig:chr22}
\end{figure}

Based on the cluster size, I can define the band range $N_b$ as $N_b=r_t \times N_r$ with $r_t$ the band ratio parameter. To avoid confusion, the band ratio $r_t$ is linearly related to sequence scale $r_b$ used in the previous sections by $r_t=r_b\frac{N}{N_c}$. Different band ratios represent structures in different scales, thus results in different clustering. Computationally, in my SeqMM, the clustering is done by using K-means on the eigenvectors from the spectral graph models. 
To explore the clustering properties with the band ratios, I consider the Hi-C data from IMR90 cell chromosome 22 with resolution 100kb. I systematically change the band ratios from 1.0, 2.0, 3.0, 5.0, 10.0, 20.0, 30.0 to 40.0.  The corresponding TAD results are illustrated in Figure \ref{fig:chr22}. Since the data size of this Hi-C data is $N=351$, when I choose $r_t=40.0$, the band matrix simply goes back the original one. It can be seen that the TAD regions evaluated from different scales is not consistent and vary a lot. Particularly when then band range change from $1.0$ to $3.0$, the calculated TAD regions are dramatically different. Even for highly correlated regions as indicated by the dark red color, TAD regions show no obvious consistence. However, when the band ratio is larger than 5.0, although the TADs are still quite different, their behaviors in these highly correlated regions become more and more consistent.

\begin{figure}
\begin{center}
\begin{tabular}{c}
\includegraphics[width=0.90\textwidth]{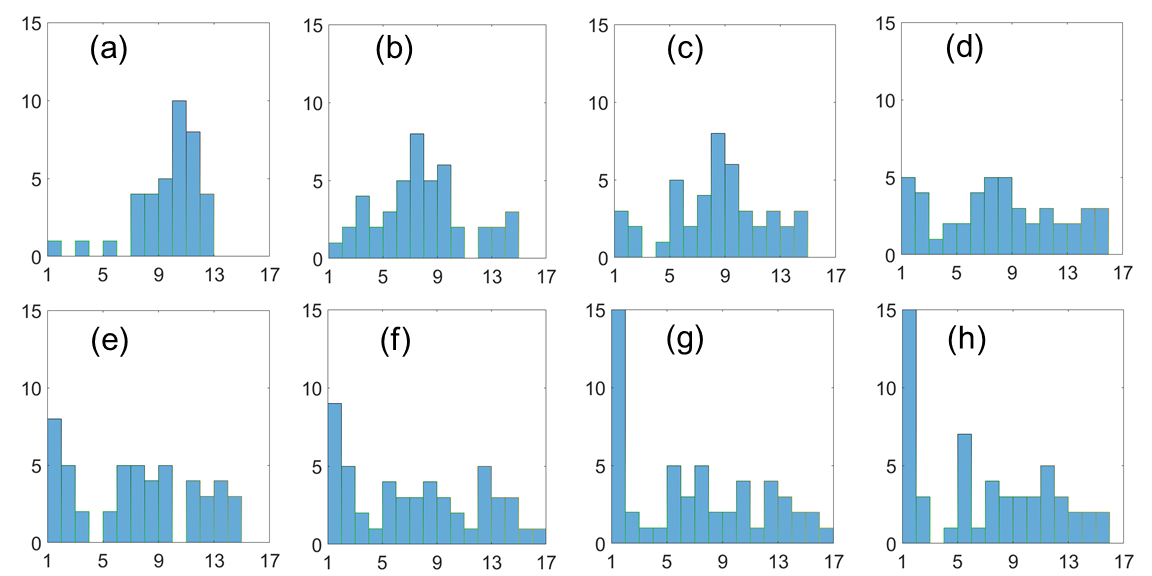}
\end{tabular}
\end{center}
\caption{ The frequency distribution of the sizes of TADs. From {\bf (a)} to {\bf (h)}, the results are derived from matrixes with  band ratios $r_t=$1.0, 2.0, 3.0, 5.0, 10.0, 20.0, 30.0 and 40.0 (the whole matrix), respectively. It can be seen that as the increase of band ratio, more and more small-sized TADs appear.
}
\label{fig:chr22_frequency}
\end{figure}

To have more quantitative understanding of this, I count the frequencies of the sizes of TADs in various band ratios. The results are demonstrated in Figure \ref{fig:chr22_frequency}. From {\bf (a)} to {\bf (h)}, the band ratios are $r_t=$1.0, 2.0, 3.0, 5.0, 10.0, 20.0, 30.0 and 40.0, respectively. The x-axis is the size of TADs in terms of loci numbers. Since the data resolution is 100kb, a single locus is made from 100 kb bases. The y-axis is the number of frequencies. It is can be see clearly that that with the increase of band ratio, small-sized TADs (at 100-200 kb) increase steadily. Moreover, when the band ratio is very small, i.e., $r_t=1.0$,  TAD size frequencies have a peak around 10 loci, which is 1 Mb in terms of bases. This peak goes to around 800 kb when band ratio is about $2.0$ to $3.0$. Further, when band ratio is larger than $5.0$, another peak will appear at 100 kb. And the sizes of the rest TADs are widely-distributed with no other obvious peak.

If one compares the TADs size frequencies in Figure \ref{fig:chr22_frequency} with the TAD regions Figure \ref{fig:chr22}, one can see that at very local scale, i.e., $r_t=1.0$, TADs are almost of the same sizes and TAD boundaries are nearly equally distributive along the diagonal regions. This means that in my localized SeqMM, the sequential information is well-preserved but Euclidean distance information within the Hi-C data is not fully utilized. When the band ratio goes to $2.0$ and $3.0$, the sizes of TAD blocks begin to diversify. Extremely small-sized TAD at 100 to 200kb and very large-sized TAD at 1300-1400 kb begin to appear. At the same time, TAD boundaries change a lot with the band ratio. No obviously consistent TAD boundaries is spotted even in regions with large contact frequencies. Mathematically, this shows that Euclidean distance information is gradually introduced and dramatically influences the clustering results. Further, when band ratio is equal to or larger than $5.0$, I begin to observe consistent TADs, particularly in the large-contact-frequency regions. More and More small-sized TADs appear, which are able to characterize the structure details of regions with inconsistent contact frequencies. Mathematically, this means the influence of the Euclidean distance information on the clustering comes to a peak at around 5.0, and gradually dwindles when band ratio passed 10.0. Further enlargement of the band ratio only introduces more long-range interactions, which slightly modify the TAD boundaries and result in more and more extremely small-sized regions. Geometrically, these small-sized regions represent the contact between two long-ranged genomic segments, similar to the situation indicated in Figure \ref{fig:illustration}. Since TADs are localized contiguous regions, these small-sized components should not be recognized as TADs. State differently, TADs should be defined not by global scaled SeqMM, instead it is better characterized by local scale SeqMM. In my SeqMM for TAD analysis, I usually use band ratio as 3.0.

\begin{figure}
\begin{center}
\begin{tabular}{c}
\includegraphics[width=0.95\textwidth]{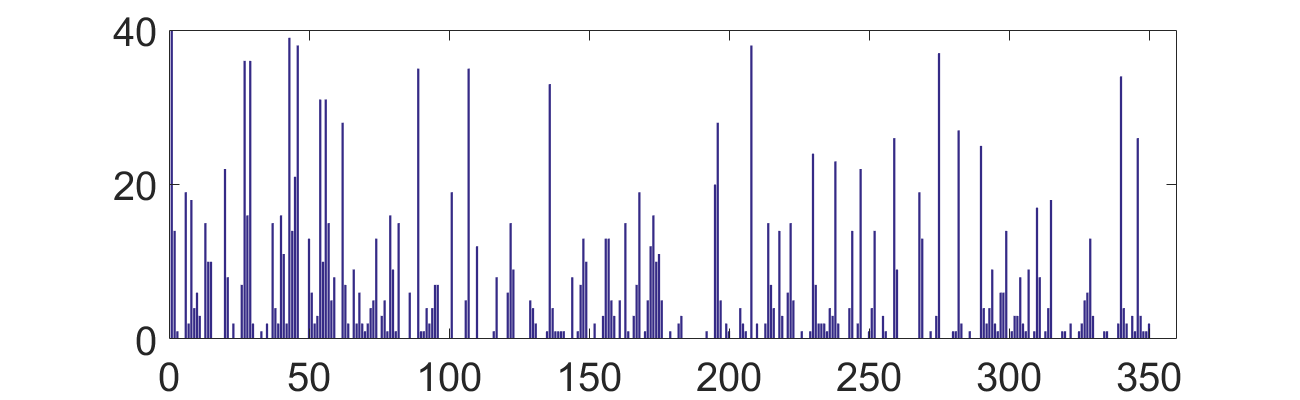}
\end{tabular}
\end{center}
\caption{The consistence of TAD boundaries in various cluster numbers. I consider the IMR90 cell chromosome 22 with resolution 100kb. The band ratio is chosen as $r_t=3.0$. I use 40 different cluster numbers ranging from 17 to 56. The frequencies of the TAD boundaries are plotted. It can be seen that the boundaries are highly concentrated around some peak values.
}
\label{fig:fre_band3}
\end{figure}

Finally, I want to show that once the band ratio (or scale parameter) is fixed at a localized scale, TAD boundaries are relatively consistent even though I use different clustering numbers. In my SeqMM, I use a fixed TAD size to define the total number of clusters in the Hi-C data. However, the size of TAD is not fixed instead it varies from 200kb to 5Mb. To explore the influence of clustering numbers on the TAD boundaries in my SeqMM, I fix the band ratio to 3.0 as in Figure \ref{fig:chr22} {\bf (c)} and vary the number of clusters from 17 to 56. I then collect all the TAD boundary values and count their frequencies. The results are illustrated in Figure \ref{fig:fre_band3}. It can be seen that instead of evenly distributed over the domain, the TAD boundaries are highly concentrated around many peaks. These high-frequency peaks are boundaries captured by my SeqMM in different clustering numbers. State differently, even though I select the clustering number based on TAD size 1Mb, the calculated boundaries in my SeqMM bear great consistence with TAD boundaries from other clustering numbers.

\begin{figure}
\begin{center}
\begin{tabular}{c}
\includegraphics[width=0.90\textwidth]{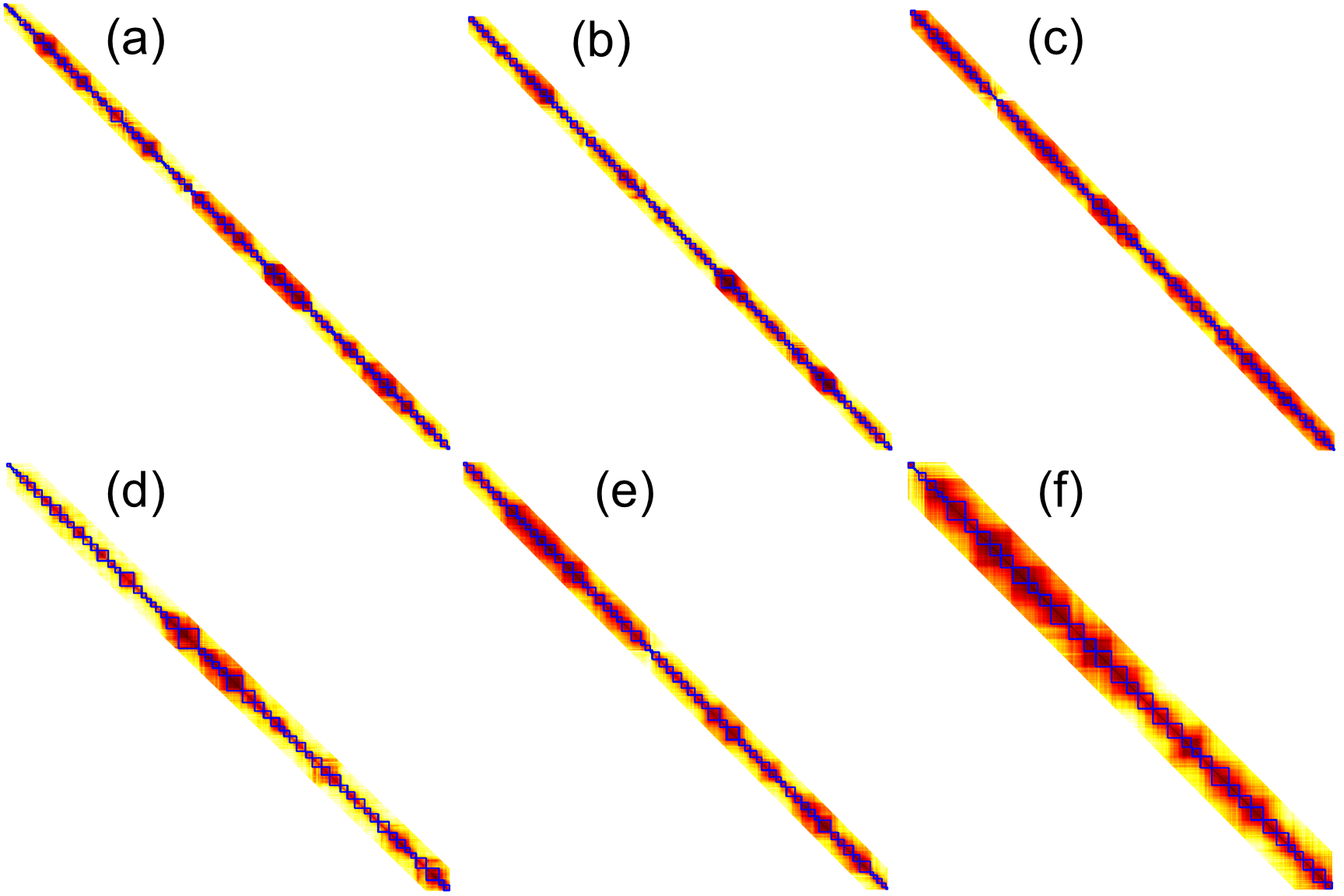}
\end{tabular}
\end{center}
\caption{ TADs of chromosome 16 to 21 for Human ES cell from Bing Ren's group. The resolution of data is 40kb. In my SeqMM, I use the band ratio $r_b=3.0$.
}
\label{fig:Bing_ren_case}
\end{figure}

To further test my SeqMM in TAD identification, I study the Hi-C data set of Human ES cell from \hyperlink{http://chromosome.sdsc.edu/mouse/hi-c/download.html}{Bing Ren's group}. The data resolution is 40kb. In my SeqMM, I use the band ratio $r_b=3.0$. For visualization, I only demonstrate the results for chromosome 16 to chromosome 21 in Figure \ref{fig:Bing_ren_case}.

\section*{Method and algorithms} \label{sec:Method_Algorithm}
In my SeqMM, the clustering is done by using K-means algorithm on the eigenvector space derived from spectral graph method. The basic procedures are as following. Firstly, I preprocess the Hi-C data to remove the insignificant rows and columns and normalize contact frequencies to a suitable weight values; Secondly, I evaluate the eigenvectors and classify them in to several clusters; Finally, I sub-decompose each cluster into several TADs until their sequence are contiguous.  A detailed discussion of basic clustering algorithms and my SeqMM algorithm is presented in the following sections.

\subsection*{Basic clustering methods} \label{sec:basic_clustering}

\paragraph{Spectral graph method} In general, a graph $G$ can be denoted as a pair $G(V,E)$, where $V= \{v_i;i=1,2,...,N \}$ denotes its set of  $N$ vertices with $N=|V|$. Here $E=\{e_i=(v_{i_1},v_{i_2});1\leq i_1 \leq N, 1\leq i_2 \leq N \}$ denotes its set of edges. The edge $e_i$ in $E$ connects a pair of vertices $v_{i_1}$ and $v_{i_2}$, and can be written as $e_i=(v_{i_1},v_{i_2})$. Then the adjacency matrix $A$ of the graph $G$ is given by \cite{ChungOverview, Mohar:1991laplacian, Mohar:1997some,von:2007tutorial}
\begin{eqnarray}\label{eq:couple_matrix28}
A_{ij}=\begin{cases} \begin{array}{ll}
	        -1 & (v_i,v_j) \in E\\
             0 & (v_i,v_j) \not \in E.\\
	      \end{array}
\end{cases}
\end{eqnarray}
Each vertex $v_i$ has a degree defined as $d_i=\sum_{i \neq j}^N A_{ij}$, i.e., the total number of edges that are connected to node $v_i$. The degree matrix $D$ can be defined as
\begin{eqnarray}\label{eq:couple_matrix27}
D_{ij}=\begin{cases} \begin{array}{ll}
	        \sum_{i \neq j}^N A_{ij} & i=j\\
            0 & i \neq j.
	      \end{array}
\end{cases}
\end{eqnarray}
With these two matrices, Laplacian matrix can be obtained as $L=D-A$. The Laplacian matrix is also known as Kirchhoff matrix or discrete Laplacian. More generally, I can define different weights on edges based on geometric, physical, chemical or biological properties or information of the data. In this way, a weighted Laplacian matrix can be represented as,
\begin{eqnarray}\label{eq:Laplacian matrix}
L_{ij}=\begin{cases} \begin{array}{ll}
            -w_{ij} & i \neq j~{\rm and} ~ (v_i,v_j) \in E\\
						-\sum_{i \neq j}^N L_{ij} &i=j\\
            0 &{ \rm otherwise.}
	      \end{array}
\end{cases}
\end{eqnarray}
A weighted graph is usually denoted as $G(V,E,W)$ with $W=\{w_{ij}; 1\leq i \leq N, 1\leq j \leq N, w_{ij}\geq 0\}$ the weighted adjacent matrix.

The Laplacian matrix has several basic properties\cite{ChungOverview,von:2007tutorial}. Firstly, it is a symmetric and semi-positive definite, thus all its eigenvalues are non-negative. Secondly, the sum of every row and every column equal to zero. In consequence, the smallest eigenvalue and the corresponding eigenvector is a unit vector. Thirdly, the rank of the Laplacian matrix is $N-N_c$ with $N_c$ the number of connected components. Finally, its second smallest eigenvalue is known as the algebraic connectivity, or Fiedler value, which represents the general topological connectivity of the graph. Moreover, the corresponding eigenvector gives an optimized classification of graph into two domains\cite{ChungOverview,von:2007tutorial}.

\paragraph{K-means method} \label{sec:K-means}
K-means clustering is a very popular method for cluster analysis in data mining. The essential idea of K-means is to minimize the within-cluster variance. To be more specific, if I want to divide a d-dimensional data $X=({\bf x_1}, {\bf x_2},...,{\bf x_n})$ into sets $S=\{\bf s_1, s_2, ..., s_k\}$, I need to find the solution to the following equation
\begin{eqnarray}\label{eq:k-means}
\underset{S}{\arg \min}\sum_i^k \sum_{{\bf x} \in {\bf s_i}} \parallel {\bf x} -E ({\bf s_i})\parallel^2=\underset{S}{\arg \min}\sum_i^k |{\bf s_i}| {\rm Var}({\bf s_i}).
\end{eqnarray}
Here $|{\bf s_i}|$, $E ({\bf s_i})$ and ${\rm Var}({\bf s_i})$ are the size, mean and variance of cluster ${\bf s_i}$, respectively.
To optimize the above objective. Fmy steps are usually employed. First, I initialize the data by setting $k$ seed values usually randomly. Then, I classify the data into $k$ cluster based on the distance between its element to the nearest seed value. After that, I can compute new seed values as the centroids, i.e., mean values, of the clusters. Finally, I repeat the above two steps until the centroids are stabilized.

\subsection*{SeqMM clustering algorithm} \label{sec:Graph}
\paragraph{Data preprocessing}
The Hi-C data, two steps are used. Firstly, I remove all the rows and columns that are summarized to zero. Secondly, I rescale the Hi-C data using the logarithm function. If I denote the original Hi-C data as,
$$M_0=\{ m_{ij}, i=1,2,...,N_0; j=1,2,...,N_0\},$$
I can find the rows and columns that need to removed,
$$I_0=\{ i \mid \sum_{j}^{N_0}m_{ij}=0 \}.$$
After the removal of these rows and columns, I have the new matrix,
$$M'=\{ m'_{ij}, i=1,2,...,N; j=1,2,...,N\}.$$
Further, I can rescale the contact frequencies to weight values by a non-negative monotonically increasing kernel,
$$W=\{ w_{ij}= f(m'_{ij}), i=1,2,...,N; j=1,2,...,N\}.$$
For example, I can use function $f(x)=log(1+x)$, so that all matrix values are non-negative.

\paragraph{Sequence-based Laplacian matrix}
The processed Hi-C data can be vieId as a weight matrix. In this way, my sequence-based Laplacian matrix can be easily constructed as following,
\begin{eqnarray}\label{eq:Laplacian matrix}
L_{ij}=\begin{cases} \begin{array}{ll}
            -w_{ij} &  |i-j|>0 ~{\rm and}~ |i-j|\leqslant N_b\\
             0 &  |j-i|> N_b\\
             -\sum_{i \neq j}^N L_{ij} &i=j
	      \end{array}
\end{cases}.
\end{eqnarray}
The parameter $N_b$ is band range of the matrix.
Based on this weighted Laplacian matrix, I then perform the eigenvalue decomposition. The resulting eigenvalues $\lambda_1<\lambda_2<...<\lambda_N$ and corresponding eigenvectors $\{{\bf u}_1, {\bf u}_2,...,{\bf u}_N\}$ can be used to study chromosome connectivity and hierarchial structure. From the definition of the Laplacian matrix, I can find that $\lambda_1=0$.

Since I want to decompose the structure into $N_c$ cluster, I choose the first $N_c$ eigenvectors to form an point cloud data set with dimension $N_c\times N$, and use K-means to cluster this data into $N_c$ clusters. For each cluster, if all loci are sequentially contiguous, I will treat this cluster as an individual TAD. However, if there is an discontinuity in their sequence, I will subdivide the cluster into subclusters along the discontinuous ones until finally all loci in each subcluster are sequential contiguous. Therefore, my finally number of TADs are usually larger than $N_c$.

My algorithm can be summarized as following:

\begin{algorithm}
\caption{SeqMM algorithm}\label{AFRI}
{\bf Pre-processing}: Remove all rows and columns, that summarized together equal to zero (or smaller than a predefined range); Rescale the Hi-C contact frequencies to a suitable weight values (Default function $f(x)=log(1+x)$);\\

{\bf Step 1}: Choose a band ratio $r_t$ (default value is 3) to construct a localized Hi-C matrix with band range $N_b=r_t\times \frac{1000}{N_r}$. Here $N_r$ is the Hi-C data resolution in the unite of kilo bases; \\

{\bf Step 2}: Construct a weight Laplacian matrix and calculate the first $N_{c}$ eigenvectors. Here $N_{c}$ is the number of clusters and $N_c=N\times \frac{N_r}{1000}$; \\

{\bf Step 3}: Employ K-means algorithm on the $N_{c}$ eigenvectors to identify $N_{c}$ clusters;\\

{\bf Step 4}: Subdivide each cluster into TADs until loci in each TAD are sequentially contiguous. \\
  \end{algorithm}

\section*{Conclusion remarks}
In this paper, I discuss the clustering of biomolecular data. Biomolecules and their complexes are hierarchical structures made from a single chain or several chains. With their sequential information, biomolecular data is fundamentally different from the general point cloud data. Therefore when sequential information matters, traditional clustering methods derived from point cloud data fall short if directly applied to biomolecular data. To overcome this problem, I propose a sequence-based multiscale model for biomolecular structure analysis. I generate a series of structural matrixes by gradually and systematically removing the long-range interactions. These new data will focus on different sequential scales and I can study the corresponding properties that are interested. More interesting, clusters from different scales give different biological implications. My SeqMM is applied to Hi-C data analysis. I have found that genomic compartment can be characterized by Fiedler eigenvector from my global scale model. Further, I study TADs with local scale models. I find that when sequence scale is small, small change of sequence scale will result in TAD boundaries that are dramatically different. No obvious consistence can be seen even for regions with high contact frequencies. However, with the increase of the scale values, although TADs are still quite different, TAD boundaries in these high contact frequency regions become more and more consistent. Further, I find that for a fixed local scale, my method can deliver very robust TAD boundaries in different cluster numbers.

\section*{Acknowledgments}
This work was supported in part by NTU SUG-M4081842.110 and MOE AcRF Tier 1 M401110000.

\vspace{0.6cm}


\end{document}